\documentclass[a4paper]{jpconf}
\usepackage{graphicx}
\def\be{\begin{equation}}
\def\ee{\end{equation}}
\def\bea{\begin{eqnarray}}
\def\eea{\end{eqnarray}}
\begin{document}
\title{Discrete symmetries and models of flavor mixing
}

\author{Alexei Yu Smirnov}

\address{International Centre for Theoretical Physics, Strada Costiera 
11, 34014 Trieste, Italy}

\ead{smirnov@ictp.it}

\begin{abstract}
%%%%%%%%%%%%%%%%%%%%%%%%%%%%%%%%%%%%%%%

Evidences of a  discrete symmetry behind
the pattern of lepton mixing are analyzed. 
The program of ``symmetry building'' is outlined.  
Generic features and problems of realization of this program 
in consistent gauge models are formulated.  
The key issues include the flavor symmetry breaking, connection of 
mixing and masses, {\it ad hoc} prescription of flavor 
charges, ``missing'' representations,  
existence of new particles, possible accidental character 
of the TBM mixing. Various ways are considered to extend the 
leptonic symmetries to the quark sector and 
to reconcile them with Grand Unification.
In this connection the quark-lepton complementarity could be  
a viable alternative to TBM.   
Observational consequences of the symmetries and future
experimental tests of their existence are discussed. 

\end{abstract}
%%%%%%%%%%%%%%%%%%%%%%%%%%%%%%%%%%%%%%%%%%%%%%%%%%%%%%%

\section{Flavor of flavor models}
%%%%%%%%%%%%%%%%%%%%%%%%%%%%%%%%%%%%%%%%%%%%%%%%

Certain features of the leptonic mixing can be considered as an evidence of 
discrete symmetry. Many different realizations 
exist \cite{reviews}. However, in spite of a various interesting developments,  
no convincing model based on discrete symmetries 
has been proposed so far. The models require new extended structures,  
many assumptions, {\it ad hoc} assignments of charges and selection of the group 
representations for multiplets. 
Additional auxiliary symmetries are needed which 
sometime even more powerful the original one. 
There are no natural and simple extensions of the leptonic symmetries 
to the  quark sector.  In most of the models no connection 
between mixing  and masses exists and  different physics (symmetry) 
is responsible for the mass hierarchies.

So, what to do? Try further using the same 
context~\footnote{In attempt to further pursue 
this approach and to check whether something interesting 
is overlooked, systematic scanning (including computerized one) of all 
possible models within certain framework has been performed~\cite{scan}.}.   
Be less ambitious and explain only some features (e.g., dominant 
structures of the mass matrices) using the symmetry? 
Or modify context (some important elements can be still missed). 
Apply symmetry differently?  

For illustration of these statements, 
several recently proposed models will be  
discussed. Generic features of the whole approach and its challenges 
are formulated. For details of specific models see talks~\cite{others} at this conference.

The paper\footnote{Talk given at the Symposium on 
``DISCRETE 2010'', 6 - 11 December 2010, La Sapienza, Rome, Italy}
 is organized as follows.  
In sect. 2 the experimental evidences in favor of discrete symmetries 
will be presented.  Sect. 3 is devoted to the Tri-Bimaximal (TBM) 
mixing and ``symmetry building''. 
In sect. 4 attempts to extend the leptonic symmetry to the quark sector are discussed. 
Sect.  5 is devoted to alternatives to the TBM approach, in particular,  to 
the quark-lepton complementarity (QLC). Sect. 6 outlines perspectives in the field, 
and conclusions are given in Sect. 7.  

\section{Data and evidences}
%%%%%%%%%%%%%%%%%%%%%%%%%%%%%%%%

\begin{figure}[h]
\includegraphics[width=21pc]{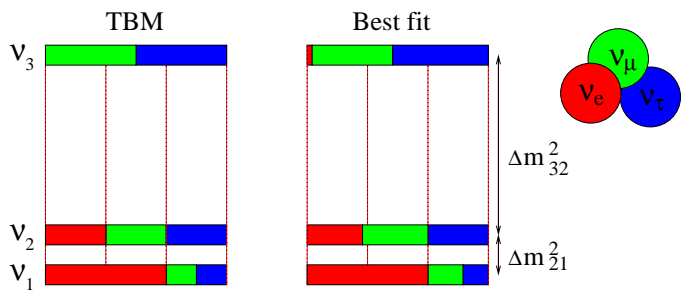}\hspace{2pc}%
\begin{minipage}
[b]{13pc}\caption{\label{fig:spec}
Mass and flavor spectrum of neutrinos in the case of TBM mixing (left) and best fit 
experimental values (right). The lengths of the colored parts are proportional to 
the moduli squared of the mixing matrix elements,  $|U_{\alpha 
i}|$. }
\end{minipage}
\end{figure}

The origin of excitement is  the neutrino mass  
and flavor spectrum shown in fig.~\ref{fig:spec}. 
Regularities of the flavors distribution are obvious:  
$\nu_\mu$ and $\nu_\tau$ flavors share $\nu_3$  equally 
(bi-maximal mixing) and  $\nu_e$ is absent in $\nu_3$, 
all three flavors are presented 
with the same weight in $\nu_2$ (trimaximal mixing). This can be formalized 
as invariance of the picture with respect to the following 
transformations: 
(1)  permutation of $\nu_\mu$ and $\nu_\tau$ flavors in all three mass states, 
which also implies zero 1-3 mixing; (2)  dilatation of the $\nu_e-$flavor part by factor 2 and 
shrinking by 2 the rest of the state $\nu_2$, 
then inverse operation in $\nu_1$: shrinking  by factor 2 the $\nu_e-$ part, and 
dilatation of the rest, and finally,  permutation of 
states $\nu_1$ and $\nu_2$. These transformations are the basis of possible discrete symmetry. 
%%
%%%%%%%%%%%%%%%%%%%%%%%%%%%%%ffff1%%%%%%%%%%%%%%%%%%%%%%%%%%%%%

%%%%%%%%%%%%%%%%%%%%%%%%%%%%%%%%%%%%%%%%%%%%%%%%%%%%%%%%%%%%%%%
%%

The regularities are described by the TBM mixing matrix~\cite{tbm}: 
\bea
U_{TBM} =  U_{23}(\pi/4) U_{12}({\rm arcsin}(1/\sqrt{3})) = \left(
\begin{array}{ccc}
            \sqrt\frac{2}{3}  & \frac{1}{\sqrt3} & 0 \\
            -\frac{1}{\sqrt6} & \frac{1}{\sqrt3} & -\frac{1}{\sqrt2} \\
            -\frac{1}{\sqrt6} & \frac{1}{\sqrt3} & \frac{1}{\sqrt2}\\
\end{array}
\right).
\label{tbmmixing}
\eea
If the length of boxes is 1,  the experimental data permit deviations 
from the pattern shown in the left figure by amount 0.01 - 0.05. 
The global fit of the oscillation data \cite{fit} agrees with the TBM mixing 
within $(1 - 2) \sigma$. It shows non-zero best fit value for the 1-3 mixing, 
some deviation of the 2-3  mixing  
from maximal one  with $\theta_{23} < \pi/4$,   and  
the 1-2 mixing slightly below $\sin^2 \theta_{12} = 1/3$:  
\be
\sin \theta_{13} = 0.118 ^{+0.038}_{-0.048}, ~~~
\sin^2 \theta_{23} = 0.463 ^{+ 0.071}_{-0.048}, ~~~
\sin^2 \theta_{12} = 0.321 ^{+ 0.016}_{-0.016}
\ee
(with $1\sigma$ errors). The deviations from the TBM values 
are small but robust  appearing  in the global analyses 
of different groups.  Important feature is that the 1-2 and 1-3 mixing angles 
correlate in the data,   
and therefore  should be considered simultaneously.   
The combined analyses of the solar and KamLaND results  \cite{sno} \cite{kamland} 
\cite{sk-sol} give non-zero 1-3 mixing at the level  
$\sin \theta_{13} = 0.02$,   $\sin^2 \theta_{12} < 0.33$,  
and agreement with TBM is at $95 \%$ CL..  
At the same time, the complete 
$3\nu-$ analysis of the atmospheric neutrino data from SuperKamiokande 
results in  $\sin2 \theta_{13} = 0.0 ~( < 0.04 )~ 90\%$,    
and $\sin2 \theta_{23} = 0.5~~(0.407 - 0.583)~ 90\%$ \cite{sk3}.  
That is, the best fit value of the 2-3 mixing  is  maximal in the case 
of normal mass hierarchy, in spite of the excess of  $e-$like sub-GeV events
which gave an indication for  $\theta_{23} < \pi/4$.

Several reservations are in order. 

1. Large deviations from the symmetry case are still possible
and if  correct (from the fundamental physics point of view) 
measure of mixing is $\sin \theta$,  then for 
1-3 mixing we have  $\sin \theta_{13} = (0.10 - 0.15)$ as the best fit 
value which should be compared with  $\sin \theta_{12} = 
0.55$.  

2. Mixing is not the RGE invariant,  and  
if theory is formulated at some high mass scales (e.g., GUT scale),  
one may expect significant deviations from the symmetry case when running to  
high energies. 

3. There are several indications that new light ($m_s  \sim 1$ eV$^2$) 
and almost sterile neutrino state exists which mixes substantially 
($\theta_{as} \sim 0.2$) with the active neutrinos. Since $m_s \sin^2 \theta_{as} \sim  0.04$ eV 
is bigger than the largest element of the mass matrix 
of active neutrinos (in the case of mass hierarchy), 
the presence of this state (or states) destroys the  present constructions
unless further complications and fine tunings are introduced. 

4. Existence of the 4th generation of fermions can change 
 the present results completely.

All this may lead to different line of developments. 
%%%%%%%%%%%%%%%%%%%%%%%%%%%%%

Concerning the mass hierarchy, there are several important features 
which can be relevant for,  but usually 
not addressed in the models under consideration:  
(1) The upper quarks have geometrical mass relation:  
$m_u m_t  =  m_c^2$; this hints that masses of all three generation should be considered 
on the same footing (and not in perturbative manner).   
(2) The down quark masses satisfy the Gatto-Sartori-Tonin relation 
$\sin \theta_C \approx  \sqrt{m_d/m_s}$ \cite{gatto} -- explicit 
relation between masses and mixing.   
(3) The charged leptons satisfy the Koide relation \cite{koide}  which  
indicates certain  symmetry, and again,  an involvement of all three generations 
of leptons.  (4) Neutrinos have the weakest mass hierarchy among  fermions which also 
shows connection between masses and mixing.

\section{TBM and symmetry building}
%%%%%%%%%%%%%%%%%%%%%%%%%%%%%%%%%%%%%%%%%%%%%%%%%%%%%%%%%%%

To a large extend  relevant symmetry can be systematically 
derived from the data \cite{reviews}, \cite{lam} and this gives the main support of the 
approach ``Mixing from discrete symmetries''. 

\subsection{Deriving symmetry}
%%%%%%%%%%%%%%%%%%%%%%%%%%%%%%%%%%%%%%%%%%%%%%%%%%%%%%%

1. In assumption that neutrinos are the Majorana particles    
the TBM mass matrix is given by  $m_{TBM} = U_{TBM} m^{diag}U_{TBM}^T$, where 
$ m^{diag} \equiv  diag (m_1, m_2, m_3)$. In the flavor basis explicitly:  
\bea
m_{TBM} =
\left(
\begin{array}{ccc}
a    & b   & b \\
...  & \frac{1}{2}(a+b+c) & \frac{1}{2}(a+b-c) \\
...  & ...  & \frac{1}{2}(a+b+c) \\
  \end{array}
\right),
\label{mTBM}
\eea
\be
a = \frac{1}{3} (2m_1 + m_2), ~~~ 
b  =  \frac{1}{3}(m_2 - m_1), ~~~c = m_3.  
\ee
Immediately one observes the $S_2$ symmetry of  
the $\nu_\mu \leftrightarrow \nu_\tau$ permutations. 
This symmetry plays a crucial role:  
it is this symmetry of the dominant structure of the mass matrix 
that ensures maximal 
2-3 mixing and zero (small) 1-3 mixing which are robust features 
of the lepton mixing. It fixes two out of three 
angles, and probably, should be a starting 
point of the symmetry building. 

The TBM-symmetry can be expressed in the form of  the 
TBM-relations: 
\be
m_{e\mu}  =  m_{e \tau}, ~~~ m_{\mu \mu} =  m_{\tau \tau}, ~~~
m_{e e}  +  m_{e \mu} = m_{\mu \mu} +  m_{\mu \tau}, 
\label{third}
\ee
or  $\sum_\alpha m_{e\alpha} = \sum_\beta m_{\mu\beta}$ instead of 
the last equality. Notice that in general, fixing three mixing angles 
leads to three relations between the matrix elements. 
What is non-trivial is that  
the relations (\ref{third}) are simple and of particular type which 
corresponds to certain symmetry.  
(In general relations are complicated and no symmetry 
can be found).   

2. The TBM mass matrix (\ref{mTBM}) in the flavor basis 
is invariant under transformations
\be
V_i m_{TBM} V^T_i = m_{TBM},
\nonumber
\ee
where
\be
V_1 = S = 
\frac{1}{3} \left(
\begin{array}{lll}
 -1  & ~~ 2 & ~~ 2 \\
 ~ ... & - 1 & ~~ 2 \\
 ~ ... & ... & - 1
\end{array}
\right), ~~~
V_2 = U = 
\left(
\begin{array}{lll}
 1  & 0 & 0 \\
 ... & 0 & 1 \\
 ... & ... & 0
\end{array}
\right).
\label{v1v2}
\ee
Two transformations (\ref{v1v2}) uniquely determine the form of the TBM mass matrix. 
They do not depend on the mass spectrum, and in fact, 
can be obtained immediately from the  $U_{TBM}$. 
At the same time, diagonality of the mass matrix squared of the charged leptons, 
$m_e^{\dagger} m_e$, can be supported by 
symmetry   
\be
V_3^\dagger  (m_e^\dagger m_e) V_3  = m_e^\dagger m_e, ~~~~ 
V_3^T = diag(1, e^{i\alpha}, e^{i\beta}), 
\ee
%%with respect to transformation   
where $\alpha \neq \beta  \neq \pi k$.  

The main idea is that $V_i$ are the generating elements 
of certain discrete symmetry group $G_f$ which eventually determines mixing. 
For instance, selecting $V_3^T = T \equiv  diag(1, \omega, \omega^2)$,
where $\omega \equiv e^{i2\pi/3}$,  one finds that    
$S, T, U$ are the generating  elements of the group $S_4$. 
Furthermore, it was argued that  $S_4$ group 
is minimal structure which leads to TBM \cite{lam} (see  
discussion in \cite{grimus}).   
(Our consideration was in the flavor basis, however neither 
mixing matrix nor symmetry  depend on the basis once the change of basis is described 
by transformation $V$ such that $V V^T = I$.)

3. The flavor symmetry $G_f$ (which contains $V_i$) should be broken.  
(In fact, no exact flavor symmetry can be introduced in whole theory, see \cite{grimus2}).)
Neutrino mass matrix is not invariant under $T$, whereas 
the charge lepton mass matrix is not invariant with respect to $S$ and $U$. 
The idea is that symmetry $G_f$  is broken {\it differently and partially}  
in the sectors which generate the neutrino masses and charged lepton masses. 
Namely,  
\be 
G_f \rightarrow  {\rm breaking} \rightarrow \left\{ 
\begin{array}{lll}
 G_\nu  & (S, U) & {\rm neutrinos}\\
 G_l & (T) &  {\rm charged~~ leptons} 
\end{array}
\right. . 
\ee
The {\it residual} symmetries $G_\nu$ and $G_l$
in the neutrino  and charged lepton sectors are different. 
Clearly $G_f$ is broken completely in whole theory. 
Furthermore, the two sectors communicated in high orders,  
and therefore $G_\nu$ and $G_l$ are broken 
even in their own sectors being approximate. 
As a result, the TBM symmetry is broken and the TBM mixing gets corrections. 

Thus, the  mixing appears as a result of different ways of the flavor 
symmetry breaking in the neutrino and charge lepton sectors. 
In turn, the difference of  symmetry breaking can originate 
(1) from  different flavor assignments of 
the right handed (RH) components of $N^c$ and $l^c$,   
or/and (2) from Majorana nature of the neutrino mass terms 
(absence of $N^c$): the neutrino and the charged lepton mass terms 
(originate from $LL$ and $L l^c$ correspondingly) 
can  have different flavor properties.

If the symmetry $G_f$ is broken spontaneously, one should introduce different sets of 
the Higgs (flavon) fields for neutrinos and charged leptons.  
Then the form of the elements of residual symmetries, $S, U$ and $T$,   
determines configurations of VEV's.  

The items 1- 3 is a paradigm of the present day flavor model building.  

4. It is possible to proceed even further in some particular way.  
The TBM mass matrix can be presented as the sum of 
three singular matrices. In the limit of small $m_1$  
it becomes as 
\be
m_{TBM} \approx
A \left(
\begin{array}{l}
 1 \\
 1  \\
 1
\end{array}
\right) \times (1, 1, 1) +  
B \left(
\begin{array}{l}
 0 \\
 1  \\
 - 1
\end{array}
\right) \times (0, 1, - 1). 
\ee
This form indicates the see-saw mechanism, for which the mass terms 
may have the form  
\be
\sum_i \frac{1}{M_i} (L f_i) (L f_i)^T , 
\label{mass}
\ee
where $M_i$ are the masses of RH neutrinos  and $f_i$ are the triplets of flavon fields,
%%(or D=5 operator origins of the matrix)  
%%(ii) universal character of couplings  
%%which can be obtained as product of two flavon triplets (
%%as in (\ref{mass})) 
(see e.g. \cite{king}). 
The VEV's of the triplets should be $\langle f_2\rangle (1, 1, 1)$ and 
$\langle f_3 \rangle (0, 1, - 1)$,  which can be obtained 
in SUSY version of model.

So far so good. Problems appear when we 
start to realize this program in consistent gauge models.

\subsection{Problems}
%%%%%%%%%%%%%%%%%%%%%%%%%%%%%%%%%%%%%%%%%%%%%%%%%%%%%%%%%%%%%%%%%%
  
Generic problem originates from the fact that masses of fermions are given by  
\be
m = F(\{Y \}, \{ v \}), 
\ee   
where $F$  is certain functional which depends on the mechanism 
of neutrino mass generation. In general,  it describes several 
different contributions and includes various corrections. 
$\{ Y \}$ are the Yukawa couplings or coupling constants 
of  different operators generating masses, and 
$\{ v \}$ refer to a set of  vev's of Higgs bosons. 
$\{ Y \}$ and $\{ v \}$ follow from independent sectors: 
from the Yukawa sector and the scalar potential.  Yukawa couplings 
are determined by symmetry (at least partially), whereas VEV's are fixed by  pattern 
of symmetry violation. In general, symmetry does not control how it is broken  
and new ingredients (dynamics, symmetries) should be introduced to fix the VEV alignment.  
%%[[??? still symmetry plays a role]].  
To obtain TBM all these components should be correlated. 
%%Sectors are tuned by some auxiliary symmetries. 
One step constructions do not work. 
Essentially this means that TBM is not immediate consequence of symmetry but 
a result of interplay of different factors, and in this sense --  accidental. 

Another generic problem is related to the fact that 
symmetry should be broken differently in the neutrino 
and charged lepton Yukawa sectors. That is, different Higgs flavon multiplets 
(typically - several for each sector) should be used.  
To forbid unwanted couplings of these flavons one is
forced to introduce  additional symmetries with {\it ad hoc} charge prescription.   

\subsection{Flavons versus Flavored Higgses}
%%%%%%%%%%%%%%%%%%%%%%%%%%%%

There are two types of models with 
broken flavor symmetries: 

1. Models with flavons $f$, singlets of the 
gauge symmetry group  which have non-zero ``flavor charges''.  
In these models  
usual Higgs doublet(s) $H$ are the flavor singlets, so that   
the electroweak symmetry and flavor symmetry breakings 
are separated. 
The Yukawa couplings and mass terms are generated by the non-renormalizable  
interactions: 
\be
\frac{1}{\Lambda^{n}_f} L e^c H f^{n} . 
\label{eq:highdim}
\ee
Here $\Lambda_f$ is the scale of flavor physics which can be above 
the GUT scale. Clearly it is difficult to test such a scenario directly, 
unless $\Lambda_f$ is not far above the electroweak scale. 

Typical problem of this scenario is existence of  high dimension operators 
of the type (\ref{eq:highdim}) which contribute to masses. 
Convergence of series is weak, $\langle f \rangle/ \Lambda_f \sim 0.2 - 0.5$,  
especially if quarks (top quark) are included in consideration.    
Further complications (e.g. additional symmetries) are needed to 
control their effect.  

2. Models with flavored Higgses: 
The Higgs doublets carry flavor charges,  and usually 
large number of such doublets (which form flavor multiplets) 
should be introduced. The flavor symmetry is broken 
simultaneously with gauge symmetry at the EW scale. 
These models are testable, and in fact,  strongly restricted 
by the FCNC, anomalous magnetic moment of muon, {\it  etc.}.  
One expects to see many scalar  bosons at LHC \cite{multi}. 

\subsection{Symmetry groups}
%%%%%%%%%%%%%%%%%%%%%%%%%%%%%%%%%%%%%%%%%%%%%%%%%%%%%%%%%%%%%%%%%

The Table 1 presents the list of small groups with  
irreducible representation ${\bf 3}$,  which are used 
in the TBM model building. 
Representation ${\bf 3}$ explains existence of three generations of fermions, 
however all these groups have also singlets (and some - doublets). 
There is no explanation why (non-trivial)  singlet and doublet representations are missing.  
An alternative is  groups, like $S_3$, with non-trivial irreducible 
representation ${\bf 2}$, so that the family structure appears as ${\bf 2} + {\bf 1}$ \cite{s333}. 

In the Table 1  we give the order of group (number of elements),  
irreducible representations and products of the representations 
which contain invariants. The latter determines the flavor structure of a model.  

%%%%%%%%%%%%%%%%%%%%%%%%%%%%%%%%%%%%%%%%%%%%%%%%%%%%%
\begin{table}[h]
\caption{\label{groups}The simplest groups with irreducible representations $\bf 
3$.}
\begin{center}
\begin{tabular}{llll}
\br
group  & order & representations & invariants  \\
\mr
$A_4$ & 12 & $1,~  1',~ 1'',~ 3$ & $3\times 3,~ 1'\times  1''$ \\
$T'$  & 24 & $1,~  1',~ 1'',~ 2,~ 2',~ 2'',~ 3$ &  $3\times 3,~ 1'\times  1'',~2\times$ \\
$S_4$ & 24 & $1,~  1',~ 2,~ 3,~ 3'$ &  $3\times 3,~ 3' \times 3',~ 2\times 2, ~ 1'\times 1'$ \\
$T_7$ & 21 & $1,~  1',~  1'',~ 3,~ 3^*$ & $3 \times 3^*$, $1'\times  1''$ \\
$\Delta (27)$ & 27 & $1_1 - 1_9$, ~ $3, ~  3'$ & $3 \times 3'$, $1_2 \times 1_3$, $1_4 \times 
1_7$, $1_5 \times 1_8$, $1_6 \times 1_9$ \\ 
\br
\end{tabular}
\end{center}
\end{table}
%%%%%%%%%%%%%%%%%%%%%%%%%%%%%%%%%%%%%%%%%%%%%%%%%%%%%%%%%%

%%In consideration of specific models I will comment on the following key issues: 
%%\begin{itemize}
%%\item 
%%existence of RH neutrinos; 
%%\item 
%%randomness ad hoc charge  prescription; 
%%\item
%%origin of mixing
%%\item 
%%TBM - stable what is special about it? [[this is another point]] 
%%\item 
%%q - l symmetry 
%%\end{itemize}

In what follows we present structures of different models based on 
various discrete symmetries. The corresponding figures show explicitly {\it ad hoc} 
character of selection of the flavon multiplets and prescription 
of the flavor charges. 
An open issue is ``missing'' representations: not all possible 
low dimensional representations are used. This may create problem 
if discrete symmetry  originates from breaking of  
some gauged continuous group \cite{luhn}.  
For each model we indicate origins of mixing 
and shortcomings with  criteria based on   
existence of auxiliary symmetries, 
presence of new fields, possibility of extension 
to the quarks sector and further embedding, {\it etc.}.

\subsection{$A_4$ symmetry and a simplest $A_4$ models}
%%%%%%%%%%%%%%%%%%%%%%%%%%%%%%%%%%%%%%%%%%%%%%%%%%%%%%%%%%%%%%%

The most popular group is $A_4$: the  symmetry group of 
even permutations of 4 elements,  or symmetry of the tetrahedron~\cite{ma}. 
It has  order 12 and two generating elements $S$ and $T$ which are needed 
to realize the TBM mixing. 
The presentation of the group: 
\be
S^2 = 1, ~~~ T^3 = 1, ~~~ (ST)^3 = 1. 
\label{repa4}
\ee
The most important element, $U = A_{\mu \tau}$, is absent. 
So, one needs either to introduce this permutation symmetry 
in addition, thus extending the symmetry group, 
or obtain it as an accidental symmetry: as a result of  
particular selection of representation and the VEV alignment.  
%%In this case $A_{\mu \tau}$ turns out to be an accidental symmetry. 

The flavor structure is determined  
by properties of products of representations and invariants: 
\be
3 \times 3 = 3 + 3 + 1 + 1' + 1'', ~~~ 1' \times 1'' =  1  
\ee
(and of course, by the VEV alignment).

%%%%%%%ffff2%%%%%%%%%%%%%%%%%%%%%%%%%%%%%%%%%%%%%%%%%%%%%%%%%%%%%
\begin{figure}[h]
\includegraphics[width=22pc]{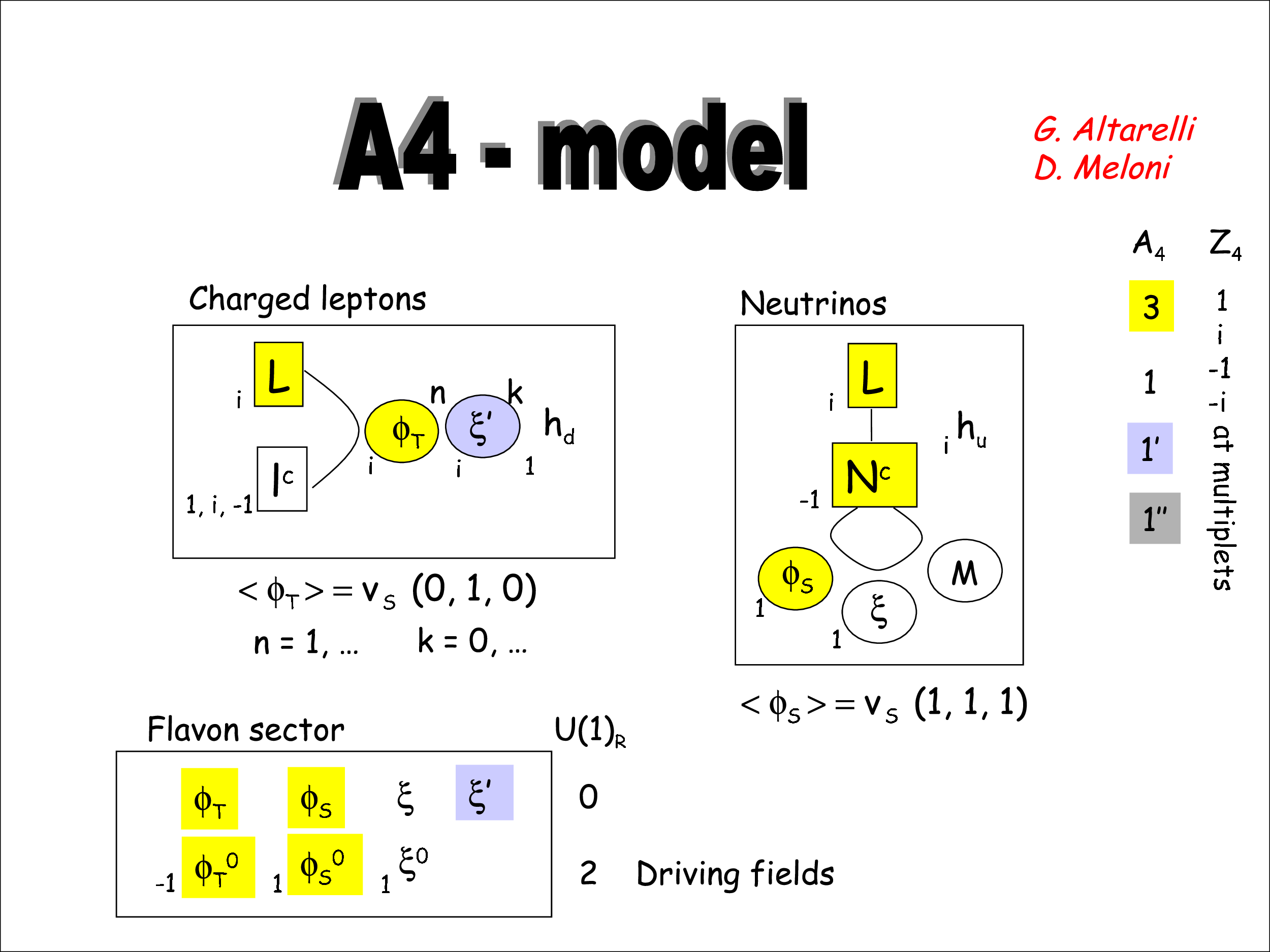}\hspace{2pc}%
\begin{minipage}
[b]{12pc}\caption{\label{a4am}
Scheme of  the lepton mass generation in the  $A_4$ model \cite{alt-mel}. 
Colors indicate representations of the flavor symmetry group ($A_4$), numbers 
at the multiplets give transformation properties with respect to  additional group ($Z_4$); 
the lines show couplings. The closed loops correspond to  self-coupling of a given multiplet, 
e.g. $N^c N^c$.}
\end{minipage}
\end{figure}
%%%%%%%%%%%%%%%%%%%%%%%%%%%%%%%%%%%%%%%%%%%%%%%%%%%%%%%%%%%%%%%%

The structure of a simplest $A4$ model for TBM mixing \cite{alt-mel} 
with charge assignment is presented in fig.~\ref{a4am}.  
The key features of the model include the following.    
There are three RH neutrinos, $N^c$. 
Four flavon multiplets participate in generation of masses. 
The lepton mixing appears due to different  
flavor assignments  of the RH neutrinos and the charged leptons:  
$N^c$ form triplet  $N^c \sim {\bf 3}$, whereas $l^c$
are all singlets, ${\bf 1}$,  of $A_4$.  
The charged leptons get Dirac masses via non-renormalizable terms,  
whereas neutrinos - via renormalizable ones.  
The $U-$symmetry is accidental: due to particular selection of the 
flavon representations and configuration of VEV's. 
The  auxiliary symmetry  $Z_4$ with {\it  ad hoc} prescribed charges is introduced;  
in particular, all $N^c$ have the same   
whereas $l^c$ have all  different $Z_4$ prescriptions. 
The vacuum alignment is achieved  by using SUSY and additional driving fields.     
The lepton mixing follows to a large extend from structure 
of mass matrix of the RH neutrinos.    
The model does not admit simple SO(10) embedding as well as the SU(5) 
embedding (if quarks also form family symmetry): 
$l^c$ are singlets of the flavor symmetry group and 
the rest fermions form triplets. 
To resolve this problem one can introduce additional  
GUT matter multiplets locating in the same multiplets 
known fermions together with new matter fields.  
$A_4$ singlet representations of $L$ are missing;  
the representation ${\bf 1''}$ is not used 
neither for matter fields nor for flavons. Apparently introduction of these missing 
multiplets will lead to new problems which could  probably be 
cured by further complications of model.

\subsection{Mixing and masses}
%%%%%%%%%%%%%%%%%%%%%%%%%%%%%%%%

In majority of the models mixing and masses are unrelated or have indirect 
relations. The latter may appear as a result of certain  
choice of the Higgs representation within a given symmetry context. 
Mixing follows from certain form invariance of the mass matrix. 
Usually, additional $U(1)$ (Froggatt-Nielsen type) or/and  discrete symmetries are used  
for explanation of the mass hierarchies.  
%%Form invariance of the mass matrix: 
Mixing is a consequence of the relations between mass matrix elements 
(like in eq.  (\ref{third})), whereas masses depend on the absolute values of the elements.  

For particular mass spectrum (set of values of masses), the mass matrix elements  are 
fixed and in some cases this may lead to more symmetric form of the mass matrix -  
to additional symmetry. Then a covering symmetry group should fix both mixing and masses. 
(E.g., one can consider the TBM mass matrix with equality of elements $a = b$,  
which gives the spectrum with normal mass hierarchy and $m_1 \approx 0$, {\it etc}.)

\section{From leptons to quarks}
%%%%%%%%%%%%%%%%%%%%%%%%%%%%%%%

Do quarks need leptonic discrete symmetries? It is not accidental 
that in the talks devoted to flavor physics in the quark  sector 
the leptonic symmetries proposed are not even mentioned. 
Although originally the first the discrete symmetries have been applied 
for flavor in the quark sector~\cite{pakvasa}.  
Presently there is no clear attempts to go ``from quarks to leptons'' 
(approach which once has been failed). The $D_{14}$ symmetry has been proposed for 
explanation of the Cabibbo angle value without extension to leptons 
\cite{claudia}. 
It is clear that the quark and lepton mixings are strongly different, 
and probably  this difference is directly related to smallness of neutrino mass.

\subsection{Extending symmetry to the quark sector}
%%%%%%%%%%%%%%%%%%%%%%%%%%%%

There are two different ways to extend the leptonic symmetries   
to the quark sector. 

1) The first possibility is to use the same representations ${\bf 3}$ and ${\bf 1}$  
for quarks  as for leptons. In the lowest order 
one can obtain  
\be
V_{CKM}^0 = I, ~~~~ U_{PMNS}^0  =  U_{TBM}. 
\ee
This difference of mixings 
can be attributed to  the Majorana nature of neutrinos.  
As a consequence of symmetry,  the Dirac 
mass matrices in the quark and lepton sectors 
are the same, both leading to zero mixing. 
(The Dirac matrices can be responcsible for the mass hierarchies 
of the charged fermions.)
The TBM follows via seesaw from certain structure of the Majorana mass 
matrix of the RH neutrinos.  Then corrections from  high order operators 
generate the CKM mixing and the deviations of lepton mixing  from the TBM form. 
Generic problem is that corrections which would explain the Cabibbo 
angle lead to too large deviations from TBM, so that additional tuning is required.   

2) Another way is to use different representations 
of the flavor symmetry group for quarks and leptons. 
In particular, one can choose groups which contain not only 
representations ${\bf 3}$ and ${\bf 1}$  but also ${\bf 2}$,  
and  assign three generations of quarks to the representations ${\bf 2}$ and  
${\bf 1}$ instead of ${\bf 3}$ in lepton sector. This implies that family symmetry 
does not exist and leaves another question: why quarks 
and leptons have different symmetry properties, that is, fundamentally different.   

\begin{figure}[h]
\includegraphics[width=22pc]{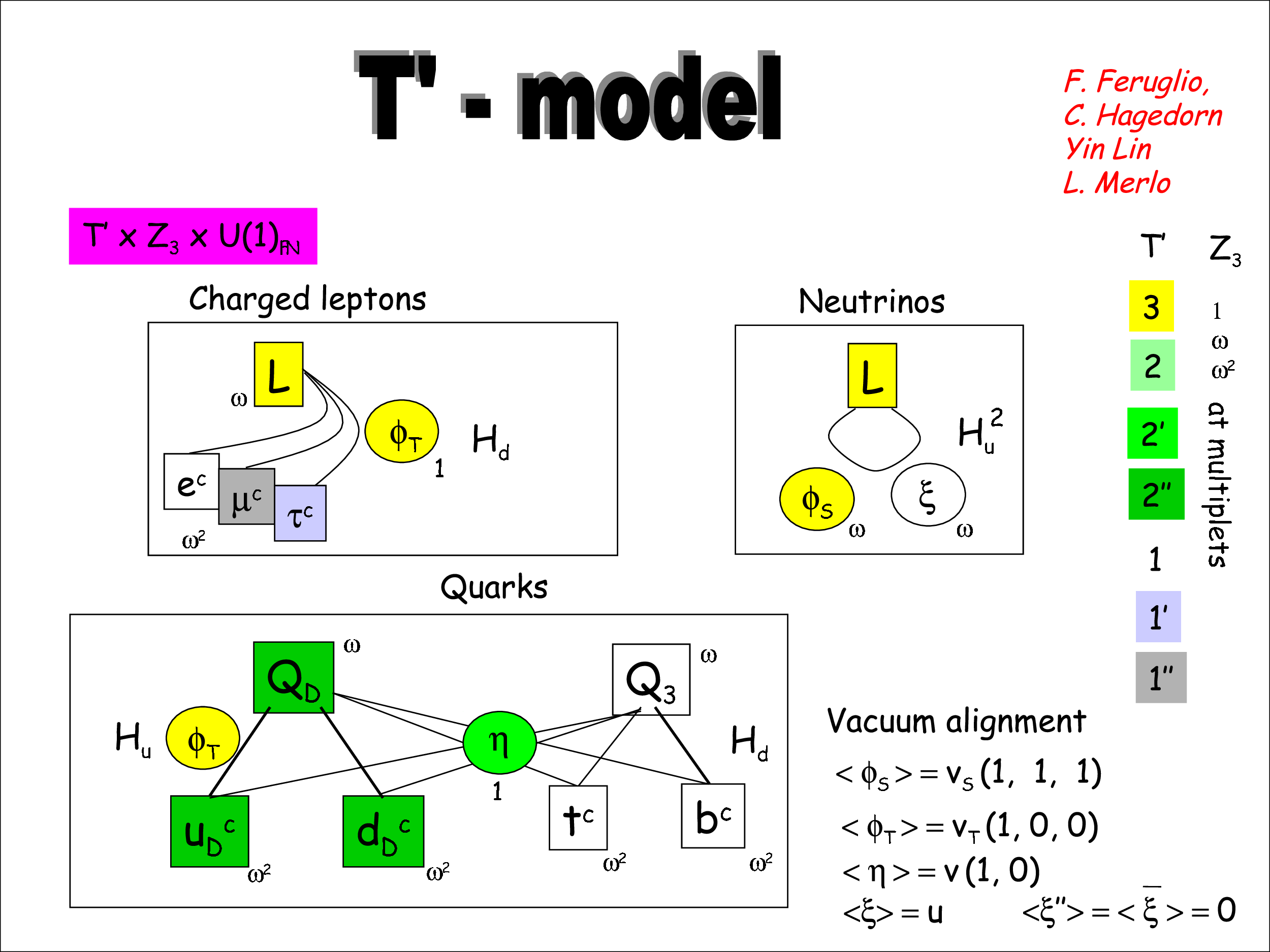}\hspace{2pc}%
\begin{minipage}
[b]{12pc}\caption{\label{tprime}
Scheme of  the mass generation in the  $T^\prime$ model~\cite{tpr-fer}. 
See the caption of fig.~\ref{a4am} for explanation.}
\end{minipage}
\end{figure}

As an example of realization of the second approach,  
consider model based on the symmetry $T^\prime$. 
The $T^\prime$ group has  order 24 being the double covering of $A_4$ or 
binary tetrahedron group. The generating elements  of this group  
 are $S$, $T$ and $R$ and presentation of the group:  
\be
T^3 = I, ~~S^2 = R, ~~~R^2 = I, ~~~ (ST)^3 = 1.  
\ee 
Here $R = 1$ for the odd-dimensional representations and 
$R = -1$ for the even-dimensional representations. Again the element $U$ is missing.  
Irreducible representations of the group include 
${\bf 1, ~1', ~1'', ~ 2,~ 2'~, 2'', ~3}$. 
The products of representations and invariants,  
\be
3 \times 3 = 1 + 1' + 1'' + 3 + 3, ~~~ 1'  \times 1'' = 1, 
\ee
coincide with those in the $A_4$ case (see Table 1).  New possibilities are related to 
existence of the doublet representations 
\be
2^a \times 3 = 2 + 2' + 2'',   
\ee
where ${\bf 2}^a = {\bf 2}, ~{\bf 2'}, ~ {\bf 2''} $ and   
\be
2 \times 2 = 1 + 3, ~~~ 1 \times 2 = 2
\label{eq:news}
\ee
with ``conservation'' of primes.  The singlet  which 
appears in (\ref{eq:news})  allows one to produce new (in comparison with $A_4$) 
flavor structures. 
The mass generation scheme of the model~\cite{tpr-fer} based on 
$T^{\prime}$  is shown in fig.~\ref{tprime}. 

%%%%%%%%%%ffff3%%%%%%%%%%%%%%%%%%%%%%%%%%%%%%%%%%%%%%%%%%%%%
%%%%%%%%%%%%%%%%%%%%%%%%%%%%%%%%%%%%%%%%%%%%%%%%%%%%%%%%%%%%%

Features of the model include the following.    
The model has an  auxiliary group $Z_3$. 
There is no  RH neutrinos, and  neutrino masses are generated by D=6 operators 
$L L H_u H_u f$, where $f = \phi_s,~ \xi$.  
The origin of mixing is the Majorana nature of neutrinos.   
In quark sector the two light generations form doublet, whereas 
the third one  is  a singlet  of $T'$.   
The RH components of charged leptons are different singlets of  $T^\prime$
but they  have the same prescription of $Z_3$.  
Four different flavon fields are introduced: two triplets and 
one doublet ${\bf 2''}$ (no ${\bf 2, 2'}$).   
Only doublets ${\bf 2''}$ are used for quarks.  
$Z_3$ prescription looks random: doublets transform with  $\omega$, 
all RH components have $\omega^2$ transformation,  
and there are various missing representations.     

\subsection{TBM and GUT's}
 %%%%%%%%%%%%%%%%%%%%%%%%%%%%%%%%

Generic problem of  many models is that the flavor prescriptions  
required for explanation of difference of mass and mixing of quarks and leptons 
prevents from their embedding into Grand Unified Theories. 
%%GUT - restricts flavor charge assignment. 
The problem  can be resolved by increasing number of matter fields and 
locating the known fermions with new ones in the same multiplets. 

One can start immediately from the GUT structure and known 
matter fields: 
\be
GUT \times G_{flavor} + {\rm new~~ elements}, 
\nonumber
\ee 
where ``new elements'' may include singlet fermions and additional Higgses 
or/and pairs of vector-like matter fields, their mixing with usual 
matter, {\it etc.}. 

\begin{figure}[h]
\includegraphics[width=22pc]{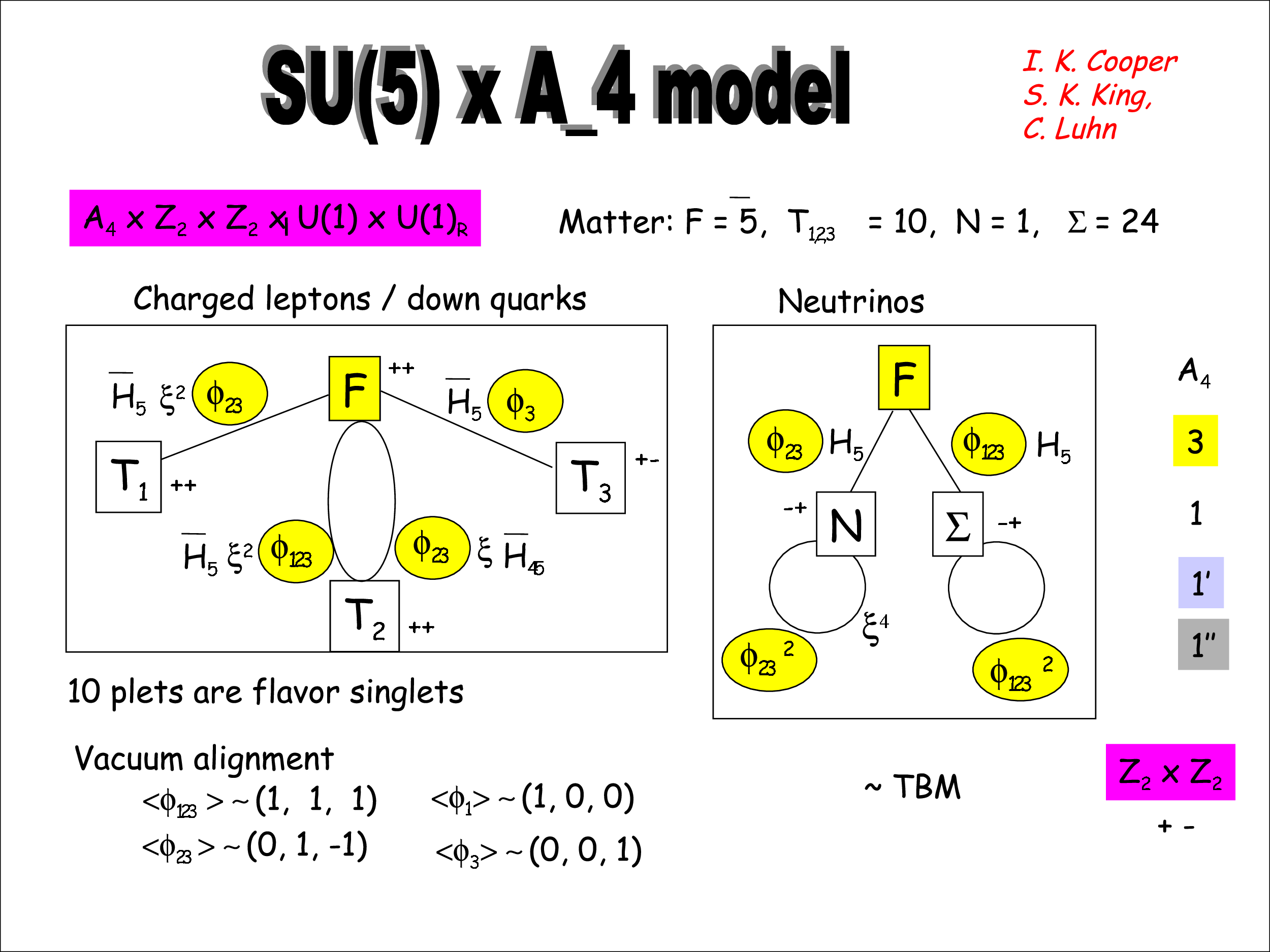}\hspace{2pc}%
\begin{minipage}
[b]{12pc}\caption{\label{su5a4}
Scheme of  generation of masses of neutrinos, charged leptons and down quarks in 
the $SU(5) \times A_4$ model \cite{cooper}. 
See the caption of fig.~\ref{a4am} for explanation.}
\end{minipage}
\end{figure}

As an example,  structure of the model based on 
$SU(5) \times A_4$ \cite{cooper} is shown in fig.~\ref{su5a4}. 
The upper quarks get masses via interactions 
$T_i T_j H_5 \{ f_{ij} \}$, 
where $\{ f_{ij} \}$ is product of certain number (from zero to 5) 
of flavon fields: e.g.,  $f_{33} = 1$, 
$f_{32} = \phi_{123} \phi_3 \xi^2$.

%%%%%%%%%%%%%%%%%%%%%%%%%%%%%%%%ffff4
%%%%%%%%%%%%%%%%%%%%%%%%%%%%%%%%%%%%%%%%%%%%%%%%%%%%%%%%%%%%%%%%
%%%%%%%%%%%%%%%%%%%%%%%%%%%%%%%%

The following remarks are in order.   
Extended auxiliary symmetry 
$Z_2 \times Z_2 \times U(1) \times U(1)$ is imposed. 
The singlet $N$  and 
$\Sigma = 24$ - adjoint representation of fermions 
are  introduced apart from usual {\bf 10}-plets and $\bar{\bf 5}-$ plets; 
neutrino masses are  generated by a combination of type the I and type III seesaw.  
Only $5-$plets form family structure, whereas $10-$plets and RH neutrinos are 
singlets of $A_4$.   
Matter fields are in  ${\bf 3, ~  1}$, but  ${\bf 1'}$, ${\bf 1''}$ are missing. 
Four different flavon multiplets  
with ``random'' $Z_2 \times Z_2$  prescriptions generate masses.  

Actually, $SU(5)$ has enough flexibility: 
three different representations ${\bf 10}, {\bf 5} , {\bf 1}$ 
allow one to write independent terms for the  upper quarks, 
for down quarks  and charged leptons and for neutrinos. 

Models based on $SO(10)$ 
with  all known fermions being in the same {\bf 16}-plets 
are more constrained.  
New elements should be added to the $SO(10) \times G_{flavor}$  
structure.  Two different ways are proposed: 
1) add singlet fermions and ${\bf 16}_H$ Higgs multiplets 
with couplings ${\bf 16} S {\bf 16}_H$ and flavons. 
In \cite{hss} $G_{flavor} = T_7$ and screening of the Dirac 
structures in the see-saw mechanism can be achieved which leads to 
independent structures of the mass matrices of neutrinos and charged fermions. 
Incomplete (partial) screening can be the origin of the TBM or bi-maximal mixings. 

2). Another way of  model building  is to introduce ${\bf 126}$  
and $\overline{\bf 126}$ plets,
and thus realize the seesaw type-II mechanism which opens up a possibility 
to obtain (to a large extend) independent flavor mixing in the 
quark and lepton sectors. Realistic model proposed 
in \cite{dutta}  and based on the symmetry $S_4 \times Z_n$ 
requires also introduction of vector-like pairs ${\bf 16}$,  
$\overline{\bf 16}$ of matter fields, additional Higgs ${\bf 10}-$plet, and flavons.  

\subsection{Is TBM accidental?}
%%%%%%%%%%%%%%%%%%%%%%%%%%%%%%%%

Experiment still allows  relatively large deviation of the mixing parameters 
from the TBM values: $\Delta \sin^2 \theta_{23} \sim  0.05$, 
$\Delta \sin^2 \theta_{12} \sim  0.02$, $\Delta \sin \theta_{13} \sim  0.15$. 
The deviations can lead to strong (maximal) 
violation of the TBM-conditions (\ref{third}), 
and consequently, to significant deviation of $m_\nu$ from the TBM form. 
For instance, instead of the first equality in (\ref{third}),  
the equality with changed sign, $m_{e\mu} \sim  - m_{e \tau}$, 
is allowed without any modification of two others.
Leading structures of the mass matrix are relatively robust, 
whereas the sub-leading structures  
can change under these corrections completely. 
It is therefore not excluded that 
the approximate TBM is  accidental being just   
an interplay of several independent  
factors (contributions) \cite{abbas}. Alternatively it can be  
a manifestation of some other symmetry 
which differs from TBM, or other structures.   
This opens up new approaches to explain the data. 

There are other possible applications of discrete symmetries. 
In the universal approach to the quark and lepton masses based on certain 
ansatz for the shape of the mass matrices  discrete symmetries are used to get 
texture zeros. In this context  the  corrected Fritzsch ansatz 
has been explored in~\cite{simoes}.  

\section{QLC and quark-lepton symmetries}
%%%%%%%%%%%%%%%%%%%%%%%%%%%%%%

The Quark-lepton complementarity (QLC)~\cite{QLC} is an alternative 
to description of the fermion mixings which is based 
on observation that 
\be
\theta_{12}^l + \kappa_{12} \theta_{12}^q \approx \pi/4, ~~~ 
\theta_{23}^l + \kappa_{23} \theta_{23}^q \approx \pi/4 , 
\ee
where $\kappa_{23}, \kappa_{12} \sim 1$,  say (0.7 - 1).  
Qualitatively, the  QLC relations mean that 

- the 2-3 leptonic mixing is close to maximal
because the 2-3 quark mixing is small; 

- the 1-2 leptonic mixing deviates from maximal one substantially because 
the 1-2 quark mixing is relatively large. 

In other words, 
\be
{\rm ``Lepton ~mixing = bi-maximal~mixing - quark~mixing''} 
\nonumber
\ee
with possible implications being: 

1. The  quark-lepton symmetry,  
which, in turn,  implies the quark-lepton unification, or GUT, or common family 
(horizontal) symmetry.  

2.  Existence of structure which produces the bi-maximal (BM) mixing.  

The structure for BM could be related to see-saw with special properties of 
the RH neutrino mass matrix.  

The Bi-maximal mixing,  $U_{bm}  = U_{23}^m U_{12}^m$,  
is characterized by maximal 1-2 and 2-3 rotations, and zero 1-3 rotation.  
There is no CP-violation. 
Possible scenario is that  in the lowest order  
\be
V_{CKM}^0  = I, ~~ U_{PMNS}^0 = U_{bm}, 
\label{eq:zorder}
\ee
and may be  $m_1 = m_2 = 0$. If the BM structure in the lepton sector 
is generated by the seesaw 
mechanism, the corrections from the Dirac mass matrix produce (i) mass splitting,  
(ii) CKM and (iii) deviation from the bi-maximal mixing. 
The situation when the deviations and CKM mixing are related
by the quark - lepton symmetry (or the same flavor symmetry for quarks and leptons) 
can be called ``strong complementarity''. 

Another possibility is Cabibbo ``haze'' \cite{haze} \cite{QLC} or 
the weak complementarity \cite{weakqlc}. 
Deviations from BM are due to some corrections which can be of the 
same order in the quark and lepton sectors but not necessarily related.  
The corrections can be of the size of the Cabibbo angle.  
%%The Cabibbo mixing parameter $\sin \theta_C  = 0.22$  
%%appears then as ``quantum''  of flavor physics. 
Possible realization is that 
the corrections are due to high order flavon interactions which generate 
simultaneously the CKM mixing and deviation from BM.  
In this case Grand Unification and  family symmetries are not necessary.

\subsection{BM-symmetry}
%%%%%%%%%%%%%%%%%%%%%%%%%%%%%%%%%%%%%%%%%%%%%%

A discrete symmetry can be behind the BM mixing as 
the lowest order structure (\ref{eq:zorder}). 
The bi-maximal mass relations 
%%(relations between elements of the neutrino mass matrix in flavor 
%%basis) 
are    
\be
m_{e\mu} = m_{e \tau}, ~~~
m_{\mu \mu} =  m_{\tau \tau}, ~~~
m_{e e} = m_{\mu \mu} + m_{\mu \tau} . 
\ee
The last equality distinguishes the bi-maximal case from  TBM (see (\ref{third})). 
The BM mass matrix, $m_{BM}$,  is invariant under transformations 
\be 
V^T_i m_{BM}V_i = m_{BM}, ~~~~~  
V_i = S_{BM}, ~~ U,   
\ee
where 
\be 
S_{BM} = 
\frac{1}{2} \left(
\begin{array}{lll}
0  & ~~ \sqrt{2} & ~~ \sqrt{2} \\
 ~ ... & - 1 & ~~ 1 \\
 ~ ... & ... & - 1
\end{array}
\right)  ~~~
\ee
with property $S_{BM}^2 = I$. 
One can select the matrix of transformation  which keeps the charged leptons mass matrix 
to be  diagonal, in the form $T_{BM} = diag (- 1, - i, i)$. In this case $T_{BM}^4 = I$, 
so that $T$ and $S_{BM}$ turn out to be the generating elements of 
$S_4$ symmetry group \cite{s4}. 

\subsection{$S_4$ symmetry and model} 
%%%%%%%%%%%%%%%%%%%%%%%%%%%%%%%%%%%%%%%%%%%%

$S_4$ has  the order 24, it is the permutation 
group of 4 elements. With  generating elements  $S_{BM}$ and  $T_{BM}$ 
it has the following presentation 
\be
S_{BM}^2 = T_{BM}^4 = (T_{BM} S_{BM})^3 = I 
\ee
(compare with (\ref{repa4})). 
%%\footnote{Another possibility is the generating elements $S$, $T$ and $U$ 
%%and the presentation
%%\be
%%T^3 = I, ~~ S^2 = I, ~~(ST)^3 = I, ~~ U^2 = I 
%%\ee
%%which shows deep connection of $S_4$ and $A_4$.}. 
It has irreducible representations 
${\bf 1, ~1', ~ 2, ~3, ~ 3'}$.
The products of representations read  
\be
3 \times 3 = 3' \times 3' = 1 + 2 + 3 + 3', ~~~~ 
3 \times 3'   = 1'  + 2 + 3 + 3' 
\nonumber
\ee
\be
1'  \times 1' = 1,  ~~~ 1' \times 2 = 2 ~~~~
2 \times 3   =  2 \times 3'   =  3 + 3' ~~~ 
2 \times 2   =  1 +  1'   + 2 , 
\ee
and the latter contains singlet, thus leading to new flavor structures. 

%%%%%%%%%%%%%%ffff5%%%%%%%%%%%%%%%%%%% %%%%%%
\begin{figure}[h]
\includegraphics[width=23pc]{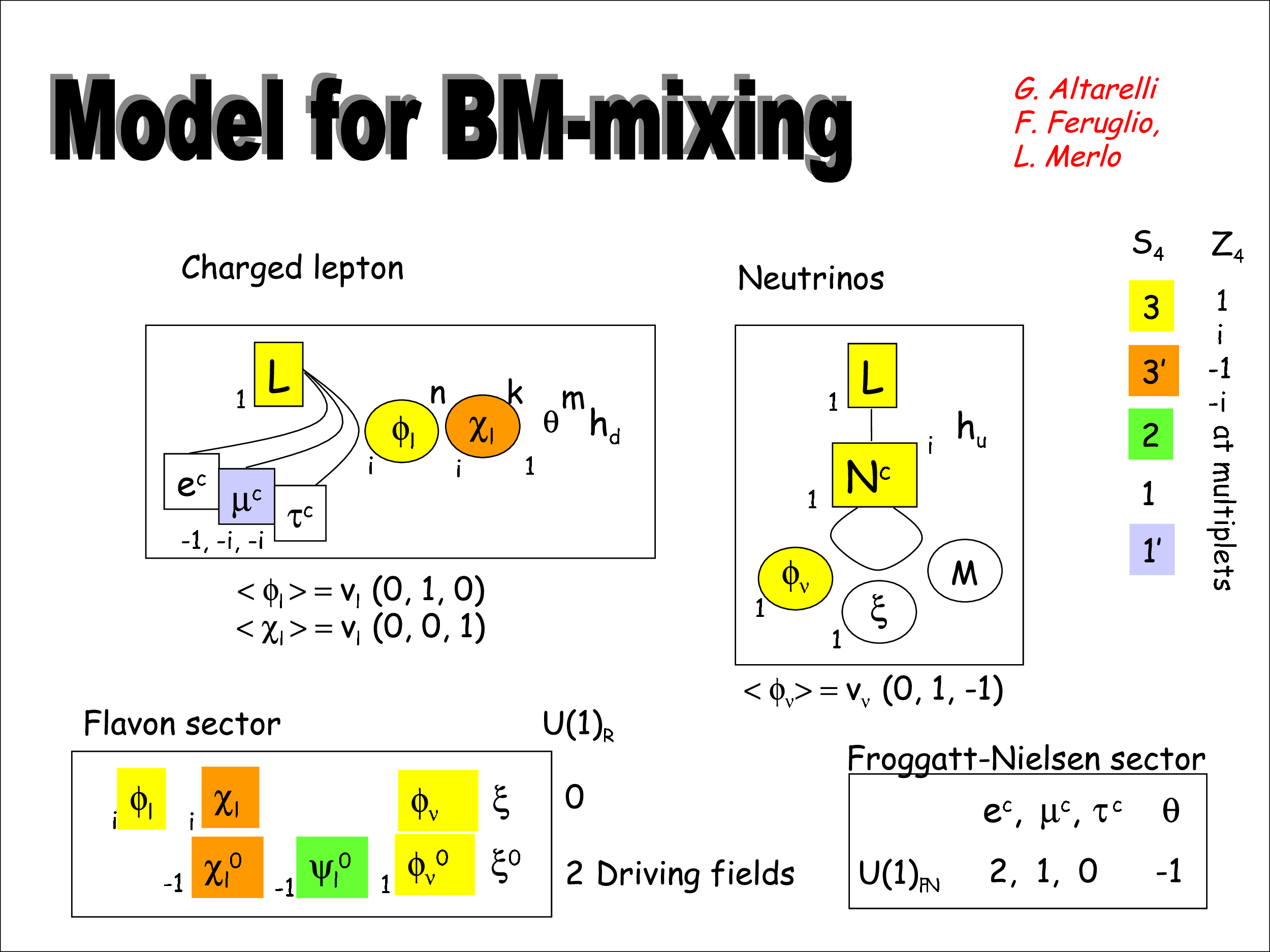}\hspace{2pc}%
\begin{minipage}
[b]{12pc}\caption{\label{bms4}
Scheme  of the $S_4$ model for the weak quark-lepton complementarity \cite{afm}. 
See the caption of fig.~\ref{a4am} for explanation.}
\end{minipage}
\end{figure}
%%%%%%%%%%%%%%%%%%%%%%%%%%%%%%%%%%%%%%%%%%%%%%%%%%%%%%%%%%%%%%%%

Structure of the model~\cite{afm} based on $S_4 \times Z_4$ and $U(1)_{FN}$ 
is shown in fig.\ref{bms4}. It resembles the structure of $A_4$ model (see Fig.~\ref{a4am}). 
The following are in order.   
Only $S_4$ representations ${\bf 3}$, ${\bf 1}$, and ${\bf 1'}$ are used 
for the  matter fields, the representations ${\bf 3'}$, ${\bf 2}$  are missing. 
Flavons are in ${\bf 3,~ 3',~ 1}$ representations,  
and ${\bf 2,~ 1'}$ are absent. 
The multiplets have {\it ad hoc} $Z_4$  prescription. 
The Froggatt-Nielsen (FN) mechanism is introduced for the mass hierarchies and only RH 
components of leptons have non-zero FN-charges.  
Deviation from BM are due to high dimension operators with flavon fields.

\section{Perspectives and tests}
%%%%%%%%%%%%%%%%%%%%%%%%

Minimal and simplest models   
which lead to the lepton mixing of the TBM or BM type from discrete symmetries, 
have been systematically explored. 
The key problem is to check the models or at least some generic features of 
the whole context. Unfortunately,  majority of the models do not give specific 
and precise predictions which can be tested. 
%% and predictions are often model dependent 
Still one expects certain connections between the low energy observables and also 
probably observables at high energy accelerators under certain additional conditions. 

In this connection several phenomenological directions should be mentioned.  
   
1. Precision measurements of the neutrino parameters.  
Determination  of  the  1-3 mixing and  the deviation of 2-3 mixing from 
maximal one are of great importance. Some models predict  $\theta_{13}$;   
discrimination of models with large and very small $\theta_{13}$ will be possible.  
Relations between $\theta_{13}$ and the deviation of $\theta_{23}$ from $\pi/2$ 
may reveal certain ways of realizations of the discrete symmetries. 

Determination of the absolute scale of neutrino masses and mass hierarchies, 
establishing possible  relations between mass ratios and mixings 
can give further insight.  

2. Double beta decay. Some models lead to certain predictions for the 
effective Majorana mass of the electron neutrino, $m_{ee}$, 
as well as  its connections to the effective electron neutrino mass, $m_e$, and sum of neutrino 
masses $\sum m_i$~\cite{bbb}. 

3. Rare decays with lepton flavor violation:  $\mu \rightarrow e + \gamma$, 
$\tau \rightarrow e + \gamma$,   $\tau \rightarrow e + \gamma$
\cite{Feruglio} \cite{mueg}. 
Equality of the rates of these decays may testify for certain class 
of models with discrete symmetries~\cite{mueg}. Interesting predictions 
for processes like $\tau^- \rightarrow e^+ \mu^- \mu^- $, 
$\tau^- \rightarrow \mu^+ e^- e^- $ are given which depend on parameter of 
violation of the discrete symmetry~\cite{Feruglio}.   

4. LHC and high energy accelerators.  
Models with flavored Higgses can be directly  tested 
in the collider  experiments~\cite{multihi}. 
Even for the SM Higgs one expects modifications 
of the decay and production rates in the presence of  
horizontal symmetries!\cite{lamhiggs}.  

5. Leptogenesis. It  is affected by discrete symmetries  
and depends on the way the symmetries are broken \cite{leptogen}. 
In some cases suppression of the leptonic asymmetry is expected.  

6. Dark matter.  Particles of the   
dark matter (e.g. in the multi-Higgs models) can be stabilized 
by some discrete symmetry which is related to the flavor symmetry~\cite{dmat}.  
For instance,   it may be a residual symmetry after breaking 
$A_4 \rightarrow Z_2$. 

7.  As already mentioned, some future discoveries can simply reject the described 
 approach or require 
its strong modification. That includes  discoveries of   
new  fermions, like  sterile neutrinos, the  4th generation of fermions, 
the right handed neutrinos or $W_R$ of the left-right symmetric models, {\it etc.}.  

\section{Conclusions}
%%%%%%%%%%%%%%%%%%%%%%%%%%%%%%%%%

In recent years, 
it has been shown that the approximate 
TBM mixing can be consistently obtained  in the context of 
gauge models with spontaneously broken flavor 
symmetries and rather extended additional structures. 
The considered examples of models show the price one should pay for realization of idea 
``mixing from discrete symmetries''.
There are two opposite  points of view on the obtained results:  

I. The features of experimental data which testify for a  symmetry behind 
lepton mixing are  actually accidental. The deviations from TBM can be significant. 
Realizations are too complicated with   
the number of assumptions being several times bigger than the number of mixing angles.   
This indicates that  alternative approaches to explanation of the data  
should be pursued. 

II. Some version of broken discrete symmetries 
give correct explanation of the data. Physics behind neutrino mass and mixing  
has rich extended structure and it leads to rich phenomenology. 
(It may happen that still some important elements of  the approach are missing. 

Preferable scenario?  
The difference of lepton and quark mixings is related 
{\it directly} to smallness of neutrino mass and probably its Majorana character. 
GUT's still look very appealing and  
there is no point to sacrifice them  in favor of 
the present models with flavor symmetries.  
The observed symmetry in the lepton mixing is related to a symmetry of  
Hidden sector at some high mass scales.  
It communicates with us via the neutrino portal --  mixing  with  neutrinos.  
No analogy of this in the quark sector exists. 
Another physics (but the same in $q-$ and $l-$ sectors) 
is involved in generation of  CKM and deviation of PMNS from  the symmetric case. 
Unfortunately, it is difficult, if possible, to check this possibility, 
but this is not the problem of Nature...

\subsection{Acknowledgments}
%%%%%%%%%%%%%%%%%%%%%%%%%%%%%%%%%%%%%%%%%%%%%%%%%%%
Author is grateful to D. Hernandez for numerous discussions of the material presented 
in this paper. 

\section*{References}
%%%%%%%%%%%%%%%%%%%%%%%%%%%%%%%%%%%%%%%%%%%%%%%%%%%%%%%%%%%%%%%%%%%%%%

\end{document}